\documentclass[pra,superscriptaddress,twocolumn,floats,10pt,floatfix,nobibnotes]{revtex4-1}

\usepackage{amsmath}
\usepackage{amsfonts}
\usepackage{amssymb}
\usepackage{graphicx}
\usepackage{color}
\usepackage[colorlinks=true,citecolor=blue,linkcolor=blue,urlcolor=black]{hyperref}
\usepackage{times}
\usepackage{amstext}
\usepackage{latexsym}
\usepackage{comment}
\usepackage{bbm}

\begin{document}

\author{Feng Hu}
\email{f.hu.121214@gmail.com}
\affiliation{Key laboratory of Specialty Fiber Optics and Optical Access Networks, Joint International Research Laboratory of Specialty Fiber Optics and Advanced Communication, Shanghai Institute for Advanced Communication and Data Science, Shanghai University, 200444 Shanghai, China}
\affiliation{Department of Physical Chemistry, University of the Basque Country UPV/EHU, Apartado 644, 48080 Bilbao, Spain}

\author{Lucas Lamata}
\affiliation{Department of Physical Chemistry, University of the Basque Country UPV/EHU, Apartado 644, 48080 Bilbao, Spain}
\affiliation{Departamento de F\'isica At\'omica, Molecular y Nuclear, Universidad de Sevilla, 41080 Sevilla, Spain}

\author{Chao Wang}
\affiliation{Key laboratory of Specialty Fiber Optics and Optical Access Networks, Joint International Research Laboratory of Specialty Fiber Optics and Advanced Communication, Shanghai Institute for Advanced Communication and Data Science, Shanghai University, 200444 Shanghai, China}

\author{Xi Chen}
\affiliation{Department of Physical Chemistry, University of the Basque Country UPV/EHU, Apartado 644, 48080 Bilbao, Spain}
\affiliation{International Center of Quantum Artificial Intelligence for Science and Technology (QuArtist) \\ and Department of Physics, Shanghai University, 200444 Shanghai, China}

\author{Enrique Solano}
\affiliation{Department of Physical Chemistry, University of the Basque Country UPV/EHU, Apartado 644, 48080 Bilbao, Spain}
\affiliation{International Center of Quantum Artificial Intelligence for Science and Technology (QuArtist) \\ and Department of Physics, Shanghai University, 200444 Shanghai, China}
\affiliation{IKERBASQUE, Basque Foundation for Science, Mar\'{i}a D\'{i}az de Haro 3, 48013 Bilbao, Spain}

\author{Mikel Sanz}
\email{mikel.sanz@ehu.eus}
\affiliation{Department of Physical Chemistry, University of the Basque Country UPV/EHU, Apartado 644, 48080 Bilbao, Spain}

\title{Quantum Advantage in Cryptography with a Low-Connectivity Quantum Annealer}

\begin{abstract}
The application in cryptography of quantum algorithms for prime factorization fostered the interest in quantum computing. However, quantum computers, and particularly quantum annealers, can also be helpful to construct secure cryptographic keys. Indeed, finding robust Boolean functions for cryptography is an important problem in sequence ciphers, block ciphers, and hash functions, among others. Due to the super-exponential size $\mathcal{O}(2^{2^n})$ of the associated space, finding $n$-variable Boolean functions with global cryptographic constraints is computationally hard. This problem has already been addressed employing generic low-connected incoherent D-Wave quantum annealers. However, the limited connectivity of the Chimera graph, together with the exponential growth in the complexity of the Boolean function design problem, limit the problem scalability. Here, we propose a special-purpose coherent quantum annealing architecture with three couplers per qubit, designed to optimally encode the bent function design problem. A coherent quantum annealer with this tree-type architecture has the potential to solve the $8$-variable bent function design problem, which is classically unsolved, with only $127$ physical qubits and $126$ couplers. This paves the way to reach useful quantum supremacy within the framework of quantum annealing for cryptographic purposes.
\end{abstract}

\maketitle

\section{Introduction}
In an era in which most of national, personal and business information is digitally stored, security of information turns out to be a major concern. This involves not only industrial, political, or diplomatic affairs, but also the information which affects to our most private circle, from finance or health care to customer patterns or political tendencies. Cryptography is, consequently, key when we try to keep the information safe from malicious parties. The security of symmetric ciphers, block ciphers, and stream ciphers mainly depends on the design of a robust non-linear combination generator, i.e. on the problem of devising $n$-variable Boolean functions satisfying multiple criteria which allow them to resist different cryptanalyses. 

Nonetheless, algebraic constructions, heuristic algorithms, and even the combination of them are still insufficient for devising cryptographically strong Boolean functions in several scenarios~\cite{carlet2010boolean, meier1988fast, courtois2003fast, carlet2002larger, tang2013highly, picek2016cryptographic, ding1991stability, zhang2017improving, carlet2002upper, pang2018construction}. Indeed, these methods are employed to design new adequate Boolean functions, but they cannot be used to generate any cryptographically strong Boolean function and, in fact, the size of the set that these techniques can generate is insignificant when compared against the total number of suitable Boolean functions~\cite{picek2013evolving, Maitra2002cons}. This leaved the brute force search in the space of $n$-variable Boolean functions as the only general method to find all cryptographically strong. However, the superexponential size of the space of $n$-variable Boolean functions $\mathcal{O}(2^{2^{n}})$ makes it intractable for classical searching~\cite{Zhang2017, Millan1997, Millan1998, Picek2016, Mariot2015}. Consequently, a new and scalable computing paradigm is required to design cryptographically strong Boolean functions.

D-Wave quantum computers are specialized quantum devices which belong to a class known as quantum annealers. The dynamics of these machines depends on an externally controlled parameter and the solution of certain optimization problem is codified in the ground state of their dynamics for certain value of the parameter. Then, starting with a Hamiltonian whose ground state is well know, the parameter is adiabatically changed until the Hamiltonian which codifies the desired solution is reached. The adiabatic theorem ensures that the dynamics ends up in the ground state. D-Wave quantum annealers have already been employed for optimization problems~\cite{JohnsonDWave}, quantum simulations~\cite{DWaveNatureSim,DWaveScienceSim}, as well as models which are classically hard to compute~\cite{DWaveNonStoq,Crisis19,Logis19}.

In Ref.~\cite{Hu2018Constr}, a general and scalable quantum-spin Ising model to design even-variable Boolean functions for their use in cryptography was experimentally implemented. In this reference, thee problem of designing Boolean functions with cryptographic criteria was mapped into the ground states of an Ising Hamiltonian which was solved in the low-connectivity D-Wave 2000Q quantum annealer. This work is remarkable since it showed the ability of the D-Wave 2000Q machine to design the Boolean functions with nonlinearity, correlation immunity, and balancedness. However, the inadequate connectivity of the D-Wave machine dramatically restricted the scalability of the embedding, since it requires an exponential number of physical qubits to codify the logical ones as $n$ grows. Consequently, it was only possible to simulate the problem for $6$-variable bent functions design and for $4$-variable Boolean functions design with high nonlinearity and resilience. The reason can be mainly attributed to the following points:

 \begin {enumerate}
\item The quantum annealer requires to codify your problem as the ground state of an Ising Hamiltonian, or equivalently a QUBO model, which restricts the pattern selection for the characterization of the original problem.

\item The topological limitation of the D-Wave hardware architecture requires the use of several physical qubits to codify a logical qubit together with its connectivity. Therefore, blocks of physical qubits must be collectively manipulated for solving complex problems, but it is often challenging and they cannot be kept simultaneously aligned. Thus, it causes additional computational errors during the quantum evolution and it may lead to a frustrated quantum annealing.

\item The size of model scales exponentially when the number of inputs and constraints, and therefore the complexity of the Ising model, increases. Hence, it is not only intractable for classical computers, but it could also be unaffordable for quantum computers. That is, the increasing demand on an exponentially growing number of qubits involved in the theoretical quantum models for characterizing special Boolean functions remains a challenge.
\end{enumerate}

Obviously, the last point is the most crucial one due to the technological limitation in the number of qubits. Nonetheless, let us remark that bent functions are the class of functions with the maximal nonlinearity and they can be characterized by a simpler model with a smaller number of qubit interactions.

In this Article, we propose a low-connectivity architecture for a quantum annealer which is scalable and optimizes the design of bent functions with large number of variables. Indeed, we prove that our tree-type architecture with no more than $3$ couplers per qubit allows a highly efficient codification of the design problem for bent functions. We estimate that a quantum annealer with  $126$ physical qubits and $127$ couplers would be able to solve the design of $8$-variable bent functions, a problem which is still classically unsolved. In our construction, two optimizations are proposed:

\begin{enumerate}
\item We show a scalable method for dimension reduction and demonstrate a feasible distributing scheme to design the bent functions. By this way, the optimal solutions can be obtained by combining the solutions given in small cases to construct the optimal solutions for large cases, and it also contributes to explore further the global property of bent functions.

\item Based on the previous point, an addapted chip architecture with a tree-type structure is engineered. The hardware is optimized for the construction of bent functions in a scalable manner employing the aforementioned dimension reduction. We estimate a much better performance when compared against D-Wave quantum annealers. Additionally, a dramatic improvement in the tradeoff between the number of qubits and the complexity of the connectivity is achieved.
\end{enumerate}
We consider that the construction of coherent quantum annealers equipped with such architecture could pave the way for achieving useful quantum supremacy in cryptography and related fields.

\section{The construction of distributed computing Ising model for bent function design}
\label{theoreticalreduction}
Generally speaking, to divide a large-scale problem into several smaller cases is effective to deal with otherwise unaffordable calculations. This is the idea behind distributed computing, where we may use many independent processors to solve a problem which, in other case, would have required much higher resources. The use of the Ising Hamiltonian for designing bent functions mainly depends on the characterization of the Walsh spectrum, which requires an exponential number of variables as the input grows linearly. Thus, our approach will be to divide the original bent function design problem into different small cases which can be addressed by a midd-size coherent quantum annealer.

\subsection{The 2-variable bent function case}
The original model of $2$-variable bent function is given by

\begin{equation}
\label{2bentfunIsing}
\begin{split}
H_\textup{non}&=\sigma _5(\sigma _1+\sigma _2+\sigma _3+\sigma _4)+\sigma _6(\sigma _1- \sigma _2+\sigma _3-\sigma _4)\\
&+ \sigma _7(\sigma _1+\sigma _2-\sigma _3-\sigma _4)+ \sigma _8(\sigma _1-\sigma _2-\sigma _3+\sigma _4).\\
\end{split}
\end{equation}
Here, the four similar terms in parentheses codify the Walsh spectrum of any $2$-variable Boolean functions represented by the qubits $\{\sigma _1,\sigma _2,\sigma _3,\sigma _4\}$ (problem qubits). In order to find bent functions, we need the ancillary qubits $\{\sigma _5,\sigma _6,\sigma _7,\sigma _8\}$, which we call controlling qubits, to guarantee that the sum $H_\textup{non}$ of the four terms is minimal. Then, we can rewrite Eq.~\eqref{2bentfunIsing} as
\begin{equation}
\label{2bentsimplify}
\begin{split}
H_\textup{non}& = (\sigma _5+\sigma _7)(\sigma _1+\sigma _2)+(\sigma _5-\sigma _7)(\sigma _3+\sigma _4)\\
&+(\sigma _6+\sigma _8)(\sigma _1-\sigma _2)+(\sigma _6-\sigma _8)(\sigma _3-\sigma _4).
\end{split}
 \end{equation}
Obviously, one of the first two terms, controlled by $\{\sigma _5, \sigma _7\}$, must be zero. Similarly, one of the last two must be zero as well. This can be used to split the problem into four conditions which are equivalent to the original problem.

\textbf{Condition 0:}

 If $\sigma _5+\sigma _7=0$ and $\sigma _6=\sigma _8$, then Eq.~\eqref{2bentsimplify} transforms into
 \begin{equation}\label{eq3}
\sigma _5(\sigma _3+\sigma _4)+\sigma _6(\sigma _1-\sigma _2).
 \end{equation}
Then, the ground state of the qubits $\{\sigma _1,\sigma _2,\sigma _3,\sigma _4\}$ can be straightforwardly derived, namely, $[1,-1,1 ,1]$, $[-1,1,1 ,1]$, $[1,-1,-1,-1]$, $[-1,1,-1 ,-1]$, each of which represents one $2$-variable bent function.

 \textbf{Condition 1:}
If $\sigma _5=\sigma _7$ and $\sigma _6+\sigma _8=0$, then Eq.~\eqref{2bentsimplify} transforms into
 \begin{equation}\label{eq4}
 \sigma _5(\sigma _1+\sigma _2)+\sigma _6(\sigma _3-\sigma _4).
\end{equation}
Similarly, the ground state of the qubits $\{\sigma _1,\sigma _2,\sigma _3,\sigma _4\}$ is given by the states $[1,1,1 ,-1]$, $[1,1,-1 ,1]$, $[-1,-1,1 ,-1]$, $[-1,-1,-1 ,1]$, each of which represents again a $2$-variable bent function.

\textbf{Condition 2:}
If $\sigma _5+\sigma _7=0$ and $\sigma _6+\sigma _8=0$, then Eq.~\eqref{2bentsimplify} transforms into
\begin{equation}
\sigma _5(\sigma _3+\sigma _4)+\sigma _6(\sigma _3-\sigma _4).
\end{equation}
In this case, the ground state only depends on $\{\sigma _3,\sigma _4\}$, thus it is not a equivalent case.

\textbf{Condition 3:}
If $\sigma _5=\sigma _7$ and $\sigma _6=\sigma _8$, then Eq.~\eqref{2bentsimplify} transforms into
\begin{equation}
\sigma _5(\sigma _1+\sigma _2)+\sigma _6(\sigma _1-\sigma _2).
\end{equation}
Similarly, the ground state only depends on $\{\sigma _1,\sigma _2\}$, and it is not equivalent to the original one.

Actually, the solutions given in Condition 0 and Condition 1 comprise all possible bent functions in the $2$-variable case. Therefore, we have shown that the $2$-variable bent function case can be divided into two equivalent smaller problems. It is noteworthy to mention that the two terms provided by Conditions 0 and 1 are independent and, hence, they can be solved separately. Consequently, only $3$ qubits and 2 couplers are sufficient to design $2$-variable bent functions, a significant dimension reduction when compared with the original model (suitable, for instance, for the D-Wave quantum annealer), which requires $8$ qubits and $32$ couplers.

Furthermore, we could consider even an additional reduction due to the symmetry under the interchange of qubits $\{\sigma _1,\sigma _2\}$ or $\{\sigma _3,\sigma _4\}$. Indeed, the ground states of the Hamiltonians in Condition 0 and Condition 1 remain the same when the labels are interchanged (it does not work for Condition 2 and Condition 3, but they do not provide any valid solution). In other words, if the solutions $\{\sigma _1,\sigma _2,\sigma _3,\sigma _4\}$ satisfies the Condition 0, then $\{\sigma _3,\sigma _4,\sigma _1,\sigma _2\}$ must satisfy the Condition 1. Consequently, it is not necessary to find both ground states, we can find every bent function by solving only one case and apply the symmetry.

On the other hand, if we divide the controlling qubits into $\{\sigma _5,\sigma _6\}$ and $\{\sigma _7,\sigma _8\}$ given the similar assumption in each group, the system can also find all the solutions in a similar way. As for the $2$-variable case, if and only if the assumptions defined on the different subgroups are symmetric, like $\sigma _5=\sigma _7$ and $\sigma _6+\sigma _8=0$, the simplified Hamiltonian can be interpreted as a small case of the $2$-variable bent function design problem.

From the point of view of the controlling qubits, the aforementioned relations among them actually characterize the distribution in the Walsh spectrum. For example, for Condition 0, the Walsh spectrum corresponding to the ground state $[1,1,1 ,-1]$ of problem qubits is given by $[2,2,2 ,-2]$, where the signs are given by the condition $\sigma _5+\sigma _7=0$ and $\sigma _6=\sigma _8$. Therefore, the assumptions about the quantum state of the controlling qubits can also help to explore the Walsh spectrum of bent functions.

To sum up, in this section, we have provided a simple demonstration of dimension reduction for the $2$-variable case, which already shows the potential for saving many quantum resources employing the symmetries between problem qubits and controlling qubits. The question is whether this reduction can be generalized to the higher variable case also employing exponentially less resources. In the following section, we will analyze the $4$-variable case aiming at finding a pattern.

\subsection{The 4-variable bent function case}
Let us generalize the approach followed int he previous section to the $4$-variable case. in the general case, there are $16$ problem qubits and $16$ controlling qubits, which may be divided into four parts as $\{\sigma _1, \sigma _2, \sigma _3, \sigma _4\}$, $\{\sigma _5, \sigma _6, \sigma _7, \sigma _8\}$, $\{\sigma _9, \sigma _{10}, \sigma _{11}, \sigma _{12} \}$, $\{\sigma _{13}, \sigma _{14}, \sigma _{15}, \sigma _{16} \}$ for the Boolean function, and $\{\sigma _{17}, \sigma _{18}, \sigma _{19}, \sigma _{20}\}$, $\{\sigma _{21}, \sigma _{22}, \sigma _{23}, \sigma _{24}\}$, $\{\sigma _{25}, \sigma _{26}, \sigma _{27}, \sigma _{28}\}$, $\{\sigma _{29}, \sigma _{30}, \sigma _{31}, \sigma _{32}\}$ for the controlling qubits.

For any group of controlling qubits $\{\sigma _i, \sigma _{i+1}, \sigma _{i+2}, \sigma _{i+3} \}$, let us denote the conditions $\sigma _i + \sigma _{i+2} =0$ and  $\sigma _{i+1} = \sigma _{i+3}$ as `0' and $\sigma _i = \sigma _{i+2}$ and  $\sigma _{i+1} + \sigma _{i+3} = 0$ as `1'. Then, four bits are sufficient to define all possible conditions in the $4$-variable case, varying from $[0, 0, 0, 0]$ to $[1, 1, 1, 1]$.

Here, we choose the constraints $[1, 1, 1, 1]$ representing the set of controlling qubits for instance, then the original Hamiltonian can be divided into two independent parts as
\begin{equation}\label{eq7}
\begin{split}
H_{\textup{4-var}}=&
\begin{bmatrix} \begin{pmatrix}
1 & 1 & 1 & 1 \\  1 & 1 & 1 & 1 \\
1 & -1 & 1 & -1 \\ 1 & -1 & 1 & -1 \\
1 & 1 & -1 & -1 \\ 1 & 1 & -1 & -1 \\
1 & -1 & -1 & 1 \\ 1 & -1 & -1 & 1
\end{pmatrix}^T \begin{pmatrix}
\sigma_1\\ \sigma_2\\ \sigma_5\\ \sigma_6\\
\sigma_9\\ \sigma_{10}\\ \sigma_{13}\\ \sigma_{14}
\end{pmatrix} \end{bmatrix}^T
\begin{pmatrix}
\sigma_{17}\\ \sigma_{21}\\ \sigma_{25}\\ \sigma_{29}
\end{pmatrix}
+\\
&\begin{bmatrix} \begin{pmatrix}
1 & 1 & 1 & 1 \\  -1 & -1 & -1 & -1 \\
1 & -1 & 1 & -1 \\ -1 & 1 & -1 & 1 \\
1 & 1 & -1 & -1 \\ -1 & -1 & 1 & 1 \\
1 & -1 & -1 & 1 \\ -1 & 1 & 1 & -1
\end{pmatrix}^T \begin{pmatrix}
\sigma_3\\ \sigma_4\\ \sigma_7\\ \sigma_8\\
\sigma_{11}\\ \sigma_{12}\\ \sigma_{15}\\ \sigma_{16}
\end{pmatrix} \end{bmatrix}^T
\begin{pmatrix}
\sigma_{18}\\ \sigma_{22}\\ \sigma_{26}\\ \sigma_{30}
\end{pmatrix}.
\end{split}
\end{equation}
These two parts can be solved independently, hence one $16 \times 16$ Walsh matrix transforms into two $8 \times 4$ Walsh matrices. Before further exploration, we use the D-Wave quantum annealer to solve them independently~\cite{Hu2018Constr}, obtaining in total $64$ $4$-variable bent function with this condition. Let us now provide a theoretical proof focusing on the first part, which we denote by $H_{\textup{half} }$. It can be rewritten as
 \begin{equation}
 \label{half}
 \begin{split}
H_{\textup{half}} =
& (\sigma_{17} + \sigma_{21})(\sigma_{1}+\sigma_{2}+\sigma_{9}+\sigma_{10}) + \\
&(\sigma_{17} - \sigma_{21}) (\sigma_{5} + \sigma_{6}+\sigma_{13}+\sigma_{14}) + \\
&(\sigma_{25}+\sigma_{29})(\sigma_{1}+\sigma_{2}-\sigma_{9}-\sigma_{10}) + \\
&(\sigma_{25} - \sigma_{29})(\sigma_{5}+\sigma_{6}-\sigma_{13}-\sigma_{14}),
 \end{split}
 \end{equation}
which following a similar analysis to the $2$-variable case, gives us the following options:

\begin{enumerate}
\item If $\sigma_{17} = \sigma_{21}$ and $\sigma_{25} + \sigma_{29} = 0$, then
\begin{equation}
\begin{split}
H_{\textup{half}} =
&2\sigma_{17} (\sigma_{1} + \sigma_{2} + \sigma_{9} + \sigma_{10}) + \\
&2\sigma_{25} (\sigma_{5} + \sigma_{6} - \sigma_{13} - \sigma_{14}).
\end{split}
\end{equation}

\item If $\sigma_{17} + \sigma_{25} = 0 $ and $\sigma_{25} = \sigma_{29} $, then
\begin{equation}
\begin{split}
H_{\textup{half}} =
&2\sigma_{17} (\sigma_{5} + \sigma_{6} + \sigma_{13} + \sigma_{14}) +  \\
&2\sigma_{25} (\sigma_{1} + \sigma_{2} - \sigma_{9} - \sigma_{10}).
\end{split}
\end{equation}
\end{enumerate}
Note that we use here $\{\sigma _i, \sigma _{i+1}\}$, $\{\sigma _{i+2}, \sigma _{i+3} \}$ as two groups for the constraints. Additionally, we can observe that both cases are again symmetric, and therefore that the solutions of any case can be derived from the solutions the other case.

Actually, a second reduction of the problem size can be achieved, since $H_{\textup{half}}$ can be divided into two smaller cases employing less qubits and couplers. Obviously, we can find $2 \times 2 = 4$ solutions for any case that we can get $4 \times 2 = 8$ solutions given that $H_{\textup{half}}$. Thus, in the case $[1, 1, 1, 1]$, we can obtain in total $8 \times 8 = 64$ bent functions of $4$ variables, which is consistent with the results obtained with the D-Wave quantum annealer~\cite{Hu2018Constr}.

Finally, we use the results obtained by the D-Wave quantum annealer~\cite{Hu2018Constr} to find the solutions for all cases from $[0, 0, 0, 0]$ to $[1, 1, 1, 1]$. Finally, we find only $8$ conditions leading to bent functions, namely, $[0, 0, 0, 0]$, $[0, 0, 1, 1]$, $[0, 1, 0, 1]$, $[0, 1, 1, 0]$, $[1, 0, 0, 1]$, $[1, 0, 1, 0]$, $[1, 1, 0, 0]$, $[1, 1, 1, 1]$. The reason is that only these $8$ conditions yield a Hamiltonian with similar symmetric properties as the described in the $2$-variable case. By taking the constraint $[1, 1, 1, 0]$ as an example, the Hamiltonian is given by
\begin{equation}
\begin{split}
H_{\textup{4-var}}=&
\begin{bmatrix} \begin{pmatrix}
1 & 1 & 1 & 1 \\  1 & 1 & 1 & -1 \\
1 & -1 & 1 & -1 \\ 1 & -1 & 1 & 1 \\
1 & 1 & -1 & -1 \\ 1 & 1 & -1 & 1 \\
1 & -1 & -1 & 1 \\ 1 & -1 & -1 & -1
\end{pmatrix}^T \begin{pmatrix}
\sigma_1\\ \sigma_2\\ \sigma_5\\ \sigma_6\\
\sigma_9\\ \sigma_{10}\\ \sigma_{13}\\ \sigma_{14}
\end{pmatrix} \end{bmatrix}^T
\begin{pmatrix}
\sigma_{17}\\ \sigma_{21}\\ \sigma_{25}\\ \sigma_{30}
\end{pmatrix}
+\\
&\begin{bmatrix} \begin{pmatrix}
1 & 1 & 1 & 1 \\  -1 & -1 & -1 & 1 \\
1 & -1 & 1 & -1 \\ -1 & 1 & -1 & -1 \\
1 & 1 & -1 & -1 \\ -1 & -1 & 1 & -1 \\
1 & -1 & -1 & 1 \\ -1 & 1 & 1 & 1
\end{pmatrix}^T \begin{pmatrix}
\sigma_3\\ \sigma_4\\ \sigma_7\\ \sigma_8\\
\sigma_{11}\\ \sigma_{12}\\ \sigma_{15}\\ \sigma_{16}
\end{pmatrix} \end{bmatrix}^T
\begin{pmatrix}
\sigma_{18}\\ \sigma_{22}\\ \sigma_{26}\\ \sigma_{29}
\end{pmatrix}.
\end{split}
\end{equation}
We can see that the position of $\sigma_{29}$ and $\sigma_{30}$ changes. Actually, controlling qubits show the distribution of Walsh spectrum, so that the two coefficients matrix exchange the last column. As a comparison with respect to condition $[1, 1, 1, 1]$, there are eight symmetric qubits groups, including $\{\sigma_{1},\sigma_{2} \}$, $\{\sigma_{3},\sigma_{4} \}$, $\{\sigma_{5},\sigma_{6}\}$, $\{\sigma_{7},\sigma_{8} \}$, $\{\sigma_{9}, \sigma_{10}\}$, $\{\sigma_{11},\sigma_{12} \}$, $\{\sigma_{13},\sigma_{14} \}$, and $\{\sigma_{15}, \sigma_{16}\}$. On the contrary, we cannot find such a symmetry for the condition $[1, 1, 1, 0]$.

On the other hand, there also exist symmetric relations between different cases. For the case $[1, 1, 0, 0]$, the symmetry cycles include $\{\sigma_{1},\sigma_{10} \}$, $\{\sigma_{2},\sigma_{9} \}$, $\{\sigma_{3},\sigma_{12}\}$, $\{\sigma_{4},\sigma_{11} \}$, $\{\sigma_{5}, \sigma_{14}\}$, $\{\sigma_{6},\sigma_{13} \}$, $\{\sigma_{7},\sigma_{16} \}$, and $\{\sigma_{8},\sigma_{15}\}$. For condition $\{1, 1, 1, 1\}$, we can find four pairs of symmetric qubits, namely, $\{\sigma_{2},\sigma_{10} \}$, $\{\sigma_{4},\sigma_{12} \}$, $\{\sigma_{6},\sigma_{14} \}$, and $\{\sigma_{8},\sigma_{16}\}$. Consequently, if the bent function $[0, 0, 0, 1, 0, 0, 1, 0, 1, 1, 0, 1, 0, 0, 0, 1]$ satisfies the condition $[1, 1, 1, 1]$, then the corresponding function $[0, 1, 0, 1, 0, 0, 1, 1, 1, 0, 0, 1, 0, 0, 0, 0]$ is also bent and satisfies the condition $[1, 1, 0, 0]$.

Therefore, as this simplified Ising model shows such a symmetric structure, it could be reduced further and used for devising bent functions. For the moment, we can get $8 \times 64 = 512$ bent functions of $4$ vriables, a subclass of all $4$-variable bent functions. That is, there must be other symmetric structures on the controlling qubits.

When growing up from 2-variable to 4-variable bent functions, we must reconsider the constraints. Here, we redefine the conditions $\sigma _i = \sigma _{i+2}$ and $\sigma _{i+1} = \sigma _{i+3}$ as `0' and $\sigma _i + \sigma _{i+2} = 0$ and $\sigma _{i+1} + \sigma _{i+3} = 0$ as `1'. Let us provide now, as an example, the Hamiltonian corresponding to the constraint $\{0, 0, 1, 1\}$
\begin{equation}
\begin{split}
H_{\textup{4-var}}=&
\begin{bmatrix} \begin{pmatrix}
1 & 1 & 1 & 1 \\  1 & -1 & 1 & -1 \\
1 & 1 & -1 & -1 \\ 1 & -1 & -1 & 1 \\
1 & 1 & 1 & 1 \\ 1 & -1 & 1 & -1 \\
1 & 1 & -1 & -1 \\ 1 & -1 & -1 & 1
\end{pmatrix}^T \begin{pmatrix}
\sigma_1\\ \sigma_2\\ \sigma_5\\ \sigma_6\\
\sigma_9\\ \sigma_{10}\\ \sigma_{13}\\ \sigma_{14}
\end{pmatrix} \end{bmatrix}^T
\begin{pmatrix}
\sigma_{17}\\ \sigma_{18}\\ \sigma_{21}\\ \sigma_{22}
\end{pmatrix}
+\\
&\begin{bmatrix} \begin{pmatrix}
1 & 1 & 1 & 1 \\  1 & -1 & 1 & -1 \\
1 & 1 & -1 & -1 \\ 1 & -1 & -1 & 1 \\
-1 & -1 & -1 & -1 \\ -1 & 1 & -1 & 1 \\
-1 & -1 & 1 & 1 \\ -1 & 1 & 1 & -1
\end{pmatrix}^T \begin{pmatrix}
\sigma_3\\ \sigma_4\\ \sigma_7\\ \sigma_8\\
\sigma_{11}\\ \sigma_{12}\\ \sigma_{15}\\ \sigma_{16}
\end{pmatrix} \end{bmatrix}^T
\begin{pmatrix}
\sigma_{25}\\ \sigma_{26}\\ \sigma_{29}\\ \sigma_{30}
\end{pmatrix}.
\end{split}
\end{equation}
We can straightforwardly observe eight pairs of qubits, namely, $\{\sigma _1, \sigma _9\}$, $\{\sigma _2, \sigma _{10}\}$, $\{\sigma _3, \sigma _{11}\}$, $\{\sigma _4, \sigma _{12}\}$, $\{\sigma _5, \sigma _{13}\}$, $\{\sigma _6, \sigma _{14}\}$, $\{\sigma _7, \sigma _{15}\}$, and $\{\sigma _8, \sigma _{16}\}$. Obviously, this condition yields 64 bent functions of $4$ variables. Finally, there are six conditions leading to bent functions, namely,$[0, 0, 1, 1]$, $[0, 1, 0, 1]$, $[0, 1, 1, 0]$, $[1, 0, 0, 1]$, $[1, 0, 1, 0]$, $[1, 1, 0, 0]$. Consequently, all the $6 \times 64 + 512 = 896$ $4$-variable bent functions can be found, all of them satisfying the symmetric structure.

The Walsh spectrum can be characterized by two kind of conditions: a) one of the pairs of controlling qubits in one group are either equal or opposite; b) both pairs of controlling qubits in one group are either equal or opposite. Independently of the condition, if there are symmetric structures, it can be reduced when finding bent functions. In general, when we generalize this approach to the $n$-variable case, during each dimension reduction, the $2^n \times 2^n$ Walsh coefficient matrix is reduced to a $2^{n-1} \times 2^{n-1}$ matrix, and the $2^n$ controlling qubits are reduced to $2^{n-2}$ until only one remains. Additionally, each reduction requires one constraint. As the input size increases, the number of constraints increases, thus we should perform several dimension reductions and then solve it using a quantum annealer.

\section{Tree-Type Quantum Annealing Architecture}
Based on the theoretical reduction on the original problem, one can prove an exponential reduction in the number of resources. However, when the variables increase, one qubit should be also connected to many qubits. Then, it is not a good choice to substantially reduce the original model. The challenge of connectivity remains in the D-Wave quantum annealer and, especially when scaling up, D-Wave cannot guarantee the equivalence of physical qubits in a chain, which produces an important constraint in the accuracy of the classical problem. In this section, we proposes a tree-type quantum annealing architecture based on the aforementioned reduction which is more suitable for our problem.

\subsection{Basic components of the quantum computing architecture}
\begin{figure}[t]
\centering
\includegraphics[width = 0.48 \textwidth]{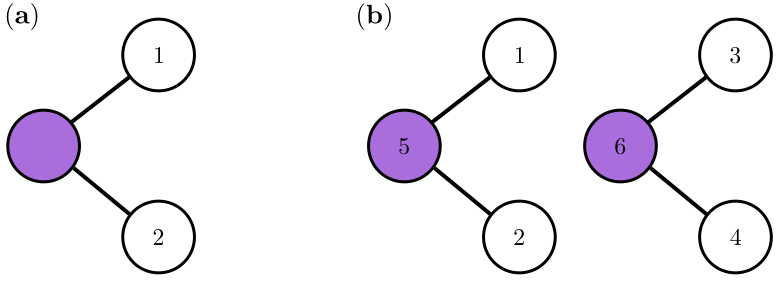}
\caption{{\bf (a)} Basic Component for Quantum Chip. The circles ``1'' and ``2'' represent two physical qubits, and the left one without number plays the role of the controlling qubit. In our case, the readout is performed in the qubits of the last layer. {\bf (b)} Architecture for devising $2$-variable bent functions obtained from Eq.~\eqref{eq3} and Eq.~\eqref{eq4}. Qubits labeled with $5$ and $6$ are controlling qubits, while qubits labeled with $1-4$ codify physical qubits.}
\label{basiccomponent}
\end{figure}

In this section, we show a simple tree-type architecture for a quantum annealing chip especially adapted to solve the bent function design problem with high accuracy and a low connectivity. Considering the high overhead of physical qubits required to codify this problem in the D-Wave quantum annealer, we aims at a better tradeoff between the number of physical qubits and the complexity of coupling connectivity. The basic component of our construction is depicted in Fig.~\ref{basiccomponent}a.

For example, let us consider the $2$-variable case, one of the cases analyzed before. The architecture codifying this problem, particularly Eq.~\eqref{eq3} and Eq.~\eqref{eq4}, is depicted in Fig.~\ref{basiccomponent}b.

Although this case can be solved in the D-Wave machine easily~\cite{Hu2018Constr}, this architecture is much simpler than the codification required by the D-Wave architecture to solve it.

\subsection{Properties and advantages of the architecture}
\begin{figure*}
\centering
\includegraphics[width = 0.8\textwidth ]{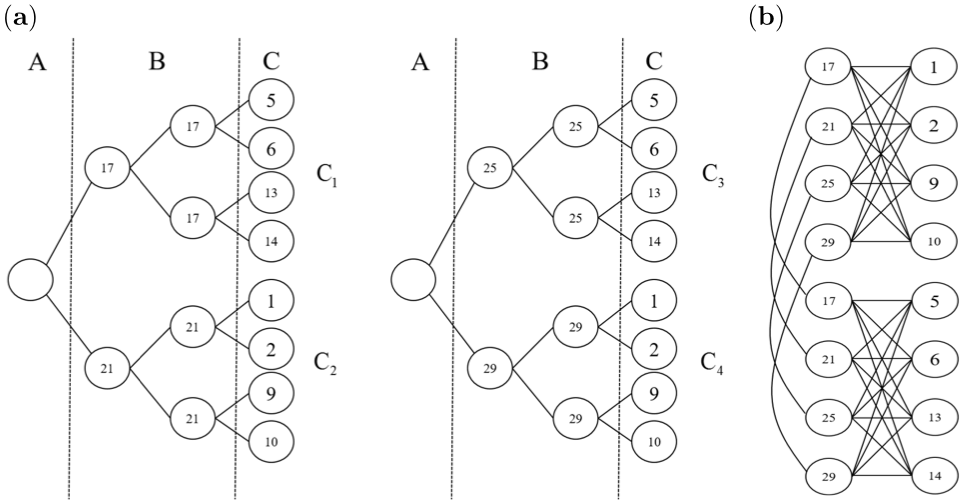}
\caption{Architecture codifying $4$-variable bent functions. {\bf (a)} Qubits in region A control the relation between different controlling qubits. In region B, it is realized the equivalence of physical qubits as a logical qubit. Region C is for final readout and the final state will be $\{C_1,C_3\}$ or $\{C_2,C_4\}$ depending on the quantum state given in Region A. {\bf (b)} Codification of $4$-variable bent function into the D-Wave Chimera graph architecture with a much higher required connectivity~\cite{Hu2018Constr}.}
\label{4varconnec}
\end{figure*}

Let us now generalize the architecture to codify the $4$-variable case under the constraint $[1,1,1,1]$. Equation~\eqref{eq7} can be rewritten as
\begin{enumerate}
\item If $\sigma_{17} = \sigma_{21}$ and $\sigma_{25} + \sigma_{29} = 0$, then
\begin{equation}
\begin{split}
&\sigma_{21} (\sigma_{1} + \sigma_{2} + \sigma_{9} + \sigma_{10}) + \\
& \sigma_{25} (\sigma_{5} + \sigma_{6} - \sigma_{13} - \sigma_{14}).
\end{split}
\end{equation}
\item If $\sigma_{17} + \sigma_{25} = 0 $ and $\sigma_{25} = \sigma_{29} $, then
\begin{equation}
\begin{split}
&2\sigma_{17} (\sigma_{5} + \sigma_{6} + \sigma_{13} + \sigma_{14}) +  \\
&2\sigma_{29} (\sigma_{1} + \sigma_{2} - \sigma_{9} - \sigma_{10}).
\end{split}
\end{equation}
\end{enumerate}
Successive reductions help us to find a tree-type architecture in which each controlling qubit connect to only {\it four} physical qubits, as shown in Fig~\ref{4varconnec}a.
Based on this, the graph can be divided into two independent parts, reading only $8$ values each time from it. Additionally, the equality of $4$ physical qubits which represent the same logical qubit is guaranteed by the tree-type architecture. If we only focus on one case, for instance $\{C_1,C_3\}$, the requirement of physical qubits and graph connectivity can be further simplified. The codification in the Chimera graph of the D-Wave quantum machine is depicted in Fig~\ref{4varconnec}b,

The proposed architecture is composed of $24$ physical qubits and $28$ couplers, while the codification in the D-Wave Chimera graph requires $16$ qubits and $36$ couplers. Although our architecture demands more physical qubits, the connectivity is substantially lower with up to three couplers per qubits. Additionally, the requirements can be further reduced in such a way that only half of the structure is sufficient to obtain the solution. We will can be analyzed it in detail in the following.

\begin{enumerate}
\item This tree-type connectivity guarantees an optimal equivalence among logical and physical qubits, as well as the stability of different spin chains, which contributes to the accuracy of the computation and leads to a complete characterization of the classical problem. However, this is still challenging for a D-Wave machine, especially when scaling up the problem. For instance, in the $8$-variable bent function design problem, the enormous overhead of physical qubits required to embed the original model in the D-Wave machine would lead to exponentially growing errors in the calculation~\cite{Hu2018Constr}.

\item From the point of view of graph connectivity, the maximal number of couplers requiered in our architecture is three, which is much smaller than in D-Wave and will consequently lead to a higher accuracy in the computation. Indeed, in the large-scale case, a controlable growth in number of physical qubits is key to guarantee the accuracy of quantum algorithm.

\item Our hardware is robust and flexible, since the accuracy of the experiment will expectably grow only by adjusting the parameters in the controlling region. This also provides an efficient manner to retrieve the information by reading out only the quantum state of the final layer, as depicted in Fig.~\ref{4varconnec}. Nonetheless, as for the D-Wave machine, too many chains involved in the quantum annealing increase the errors, which would inexorably require a classical post-processing.

\item From the point of view of experimental feasibility, our graph is completely symmetric and scalable, which also provides an straightforward manner to embed the problem with more qubits but fewer couplers. Our approach breaks the initial unaffordable problem into solving several independent smaller problems, which classifies this algorithm as a practical distributed quantum computing algorithm~\cite{BBGHKLSS13}.
\end{enumerate}

Essentially, the tree-type graph for the architecture allows us to achieve a better tradeoff among the number of qubits, the coupler strengths, and the accuracy of quantum computing algorithm. By introducing a few extra physical qubits, the complexity of the connectivity is dramatically reduced and allows us to split the problem into several independent smaller problems. All-in-all, a better ratio between physical an logical qubits is obtained, which is an effective way to realize a more robust quantum annealer.

\section{Feasibility of the Implementation}
In previous sections, we have decomposed the problem into several subproblems and described the codification in an optimal low-connectivity quantum-annealing architecture. In this section, we will analyze the feasibility of the proposal in superconducting circuits. In particular, we will focus on the resources demanded to address the 8-variable bent function design problem, the first classically unsolved case. Our tree-type architecture requieres 126 qubits and 127 couplers to codify this problem, with a coordination number between one and three, as depicted in Fig.~\ref{4varconnec}a. As a reference, let us compare these numbers with the D-Wave 2000Q quantum annealer with a connectivity given by the Chimera graph. This processor counts with 2048 functional qubits with a coordination number larger than or equal to 5, as indicated in Fig.~\ref{4varconnec}b, which sums up to 6016 couplers. Another relevant number is that it comprises 200 I/O and control lines, which means that at least 200 of the qubits can be controlled and measured directly. In 2019, D-Wave has announced the next generation of quantum processors called Pegasus \cite{pegasus} with 5000 qubits and connectivity 15. Consequently, the requirements of our architecture in terms of number of qubits and connectivity are quite below the state of the art in quantum annealers.

The main limitation of D-Wave 2000Q is the incoherence of the dynamics shown by the device. The consequence is that the sampling, and therefore the computational time, must be increased to achieve the same accuracy in the outcome. The growth of the sampling to reduce the SNR is polynomial  and depends on several factors, e.g. the temperature, decoherence ratio, among others \cite{KSBKWGO19}. There will be a threshold for the decoherence time such that, the sampling required to retrieve the result is so large, that it overcomes the classical computation time.

A possible alternative is to reduce the number of quantum resources required. Indeed, our algorithm allows to reduce the number of quantum resources by increasing the classical computational complexity. Therefore, one could consider a smaller quantum chip comprising 63 qubits by increasing approximately in a factor of 4 the classical resources required, which would still be affordable by current HPC facilities. The reference is then the Sycamore chip from Google~\cite{Aetal19}, which contains 142 (coherent) transmon qubits: 54 qubits have individual microwave and frequency controls and are individually read out (qubits) and the remaining 88 transmons are operated as adjustable couplers remaining in their ground state (couplers). This quantum processor, which is highly coherent, can be employed as a quantum annealer, as well. The resources comprising this chip are close to our requirements, but the connectivity graph is not the correct one, since it is a square lattice instead of a tree-type architecture. However, it clearly shows the feasibility of our proposal.

\section{Conclusions and Outlook}
We have proposed a scheme to reduce the superexponential dimension scaling associated to the problem of devising bent functions. This scheme allows us to split the problem into several independent smaller cases which can be efficiently embedded in a low-connectivity tree-type quantum annealing architecture. Additionally, some proofs with experiments on devising 4-variable bent functions via a D-Wave quantum computer are provided for demonstrating scalability and completeness of our approach for solving large-scale $2n$-variable bent function problem~\cite{Hu2018Constr}. Our proposal of tree-type chip proves to scale up better and be more stable allowing higher accuracy in the computation.

From the point of view of quantum supremacy, $8$-variable bent-function design is computationally hard for classical computers and a complete determination has not been achieved yet. In our approach, we may divide the $8$-variable bent function design problem into $4$ parts, each of them can be independently codified into $16$ logical controlling qubits and $64$ logical problem qubits, summing up to $80$ logical qubits in total. Then, it can be embedded into $127$ physical qubits with only $126$ couplers to solve each part, which shows a significant improvement compared to D-Wave quantum annealers, which requires $512$ physical qubits and $1456$ physicals couplers for each part~\cite{Hu2018Constr}. This would allow us to achieve useful quantum supremacy with only $127$ qubits in the framework of a distributed quantum algorithm.

\section*{Acknowledgements}
We acknowledge funding from projects QMiCS (820505) and OpenSuperQ (820363) of the EU Flagship on Quantum Technologies, as well as the EU FET Open Grant Quromorphic (828826), Spanish Government PGC2018-095113-B-I00 (MCIU/AEI/FEDER, UE) and Basque Government IT986-16, Shanghai Municipal Science and Technology Commission (18010500400 and 18ZR1415500), the National Natural Science Foundation of China (Grants 61332019, 61572304, and 61272096), and the Shanghai Program for Eastern Scholar. This work is supported by the U.S. Department of Energy, Office of Science, Office of Advanced Scientific Computing Research (ASCR) quantum algorithm teams program, under field work proposal number ERKJ333.

\end{document}